\documentclass[twocolumn,english,aps,prb,showpacs]{revtex4}
\usepackage[]{fontenc}
\usepackage[latin1]{inputenc}
\usepackage{amsmath,amssymb,babel,graphicx}
\begin{document}

\title{Magnetic droplets in a metal close to a ferromagnetic quantum critical point}

\author{Y.~L.~Loh}

\affiliation{Theory of Condensed Matter Group, Cavendish Laboratory, Department
of Physics, University of Cambridge, Madingley Road, Cambridge CB3
0HE, United Kingdom}

\author{V.~Tripathi}

\affiliation{Theory of Condensed Matter Group, Cavendish Laboratory, Department
of Physics, University of Cambridge, Madingley Road, Cambridge CB3
0HE, United Kingdom}

\author{M.~Turlakov}

\affiliation{Theory of Condensed Matter Group, Cavendish Laboratory, Department
of Physics, University of Cambridge, Madingley Road, Cambridge CB3
0HE, United Kingdom}

\begin{abstract}
Using analytical and path integral Monte Carlo methods, we study the
susceptibility $\chi_{dc}(T)$ of a spin-$S$ impurity with
$XY$ rotational symmetry embedded in a metal. 
%and interacting with the
%conduction electrons through an antiferromagnetic exchange coupling.
Close to a ferromagnetic quantum critical point, the impurity polarizes conduction electrons 
in its vicinity and forms a large magnetic droplet with moment $M\gg S.$ At not 
too low temperatures, the strongly damping paramagnon
modes of the conduction electrons suppress large quantum fluctuations
(or spin flips) of this droplet. We show that the susceptibility
follows the law $\chi_{dc}(T)=(M^{2}/T)[1-(\pi g)^{-1}\ln(gE_{0}/T)]$,
where the parameter $g\gg 1$ describes the strong damping by
conduction electrons, and $E_0$ is the bandwidth of paramagnon modes.
%, and $g$ increases as one approaches criticality. 
At exponentially
low temperatures $T\ll T_{*} \sim E_{0}\exp(-\pi g/2)$ we show that
spin flips cannot be ignored.  
In this regime we find
that $\chi_{dc}(T)\approx\chi_{dc}(0)[1-(2/3)(T/T_{*})^2]$,
where $\chi_{dc}(0)\sim M^{2}/T_{*}$ is finite and exponentially large in $g$.
We also discuss these effects in the context of the multi-channel
Kondo impurity model. 
\end{abstract}
\maketitle

%==============================================================================
\section{Introduction}

Magnetic impurities in a nearly ferromagnetic Landau Fermi-liquid
can induce large magnetic droplets by polarizing conduction electrons
in their vicinity.\cite{larkin} 
The large magnetic polarizability of the conduction electrons  
can be described in terms of low-energy collective excitations, 
paramagnons.\cite{doniach}
An impurity spin dressed by these soft
paramagnon excitations forms a magnetic droplet,  and
the size of such droplets, determined by the spatial dispersion of the paramagnons, 
can greatly exceed typical interatomic distances
in the proximity of the quantum critical point.  
The dynamics of such a droplet
are essentially determined by the paramagnon modes
which damp the orientational motion of the droplet. 
At not too low temperatures, the fluctuations of the droplet's
moment are small if the damping of the angular motion is strong.
As the temperature is lowered, due to these small fluctuations
the effective damping decreases slowly, usually as a 
power law or logarithmically (see below).
This decrease of the effective damping is reflected in the temperature
dependence of the impurity's magnetic susceptibility which increases 
at a rate slower than the Curie-Weiss law, $\chi_{dc}(T)\sim T^{-1},$ as the
temperature is decreased. 
%In this paper we address the question
%whether for a droplet with a continuous ($XY$) rotational symmetry, \emph{large} quantum fluctuations
%could become important at sufficiently low temperatures, and if that
%is the case, the effect they have on the impurity susceptibility.

The relevance of large quantum fluctuations (or spin flips) 
of these overdamped magnetic droplets, the main interest of this
paper, has been
a topic of active investigation recently. Consensus in this matter
has proved elusive. 
The various existing points of view seem to agree
that the susceptibility of large overdamped droplets at not very low
temperatures should obey $\chi_{dc}(T)\sim T^{-1+\alpha},$ where
$\alpha<1$ is non-universal. 
Millis, Morr and Schmalian\cite{millis1} found that in
the case of a magnetic defect with Ising symmetry, quantum tunneling
is suppressed altogether, and that the power-law temperature dependence
of $\chi_{dc}(T)$ extends down to $T=0~K$. Furthermore,
they indicated that their conclusion was relevant even for defects
with a continuous symmetry.\cite{comment1} Castro-Neto 
and Jones\cite{castro-neto1}, in an earlier work, analysed the same
problem with both ferromagnetic and antiferromagnetic clusters. 
In a broader context, our analysis forms a part of the 
general problem of disorder
in nearly quantum critical metals. Three decades ago, 
Griffiths,\cite{griffiths1}
and McCoy,\cite{mccoy1} predicted a non-analytic temperature dependence
of magnetization in nearly-ordered ferromagnetic Ising models with
bond disorder. Throughout the years after that, the research 
on quantum Griffiths, Kondo disorder and local criticality\cite{coleman,si}
problems remains of wide interest.

We study the magnetic susceptibility, $\chi_{dc}(T),$
of a spin-$S$ impurity with $XY$ rotational symmetry coupled
to the conduction electrons through an exchange interaction, 
\begin{align}
H_{\mbox{ex}} &= J\,\mathbf{S}\cdot c^{\dagger}(0){\pmb\sigma}
c(0),
\label{Hex}
\end{align}
 in a metal close to a ferromagnetic quantum critical point. 
%The sign  of the exchange coupling $J$ is usually antiferromagnetic,
% $(J>0),$ for typically, the moment is formed by the
% Anderson mechanism.\cite{anderson1} 
%The results we obtain below are independent of the sign of $J$.
% However the effects we study in this paper exist alongside other
% contributions, such as the usual Kondo effect, 
%that depend on the sign of $J$. 
We employ both analytical and path integral Monte Carlo (PIMC) techniques. The
two main results of our analysis are as follows. First, we show that
at not very low temperatures, due to small quantum fluctuations of the $XY$
droplet the susceptibility evolves as \begin{align*}
\chi_{dc}(T) &= \frac{M^{2}}{T}\left[1-\frac{1}{\pi g}\ln\left(\frac{gE_{0}}{T}\right)\right].\end{align*}
We shall show that the result is valid over an exponentially large
range of temperature, $gE_{0}\gg T\gg E_{0}\exp(-\pi g/2).$
Here the parameter $g\approx(Jn(\epsilon_{F}))^{2}/(1+F_{a})\gg1$ 
represents the
damping of the droplet's fluctuations, and 
$|M|\approx |J|n(\epsilon_{F})S/(1+F_{a})\gg S$
is the magnetic moment of the droplet. $F_{a}\rightarrow -1$ is the Landau
Fermi-liquid parameter denoting distance from the quantum critical
point. To leading order in $g^{-1},$ the logarithmic form and the
power law, $\chi_{dc}(T)=(T/T_{0})^{-1+1/(\pi g)},$ are the same.
However at lower temperatures, the two expressions can differ significantly.
Second, and more significantly, we show that at exponentially low
temperatures $T\ll T_{*}=E_{0}\exp(-\pi g/2),$ full $2\pi$ spin rotations
cannot be ignored. The susceptibility at $T=0$ saturates in this model, and at
 finite temperatures $T\ll T_{*}$, 
\begin{align*}
\chi_{dc}(T) &\approx \chi_{dc}(0)[1 - (2/3)(T/T_{*})^2],\end{align*}
 where $\chi_{dc}(0)\sim M^{2}/T_{*}$ is of the order of the susceptibility
at $T \sim T_{*}$. %where spin flips could be disregarded. 
The zero-temperature susceptibility $\chi_{dc}(0)$
is exponentially large in $g.$ The conclusion is that for the $XY$
magnetic defect, spin flips are important at low temperatures,
and their effect is to remove the divergence of susceptibility at
 $T=0~K$. The low-temperature behavior we obtain for the impurity susceptibility
 is similar to that seen in the usual Kondo problem\cite{anderson1,hewson} for
 $J>0$. 
The regime of proliferation of spin flips below $T<T_{*}$ is analogous
to the strong-coupling regime of the Kondo problem below the Kondo temperature.
In contrast to the usual Kondo effect where the sign of the exchange coupling between 
the impurity spin and conduction electrons, $J$, is important, our results are independent 
of the sign of $J$.  We consider here only the effects dependent on the damping
$g$ and therefore on even powers of $J$. 
In the closing section we discuss further the experimental realizations
and differences of magnetic droplet phenomena from the standard Kondo effect\cite{hewson,moriya}.
%The effect
%of odd powers of $J$ can be neglected if we assume that
%the Kondo temperature $T_K$ is much lower than the crossover temperature $T_{*}$,
%therefore requiring $Jn(\epsilon_{F}) \ll (1+F_a)^{1/3}$.
%Sufficiently close to the critical point ($g\gg1$), the
% characteristic temperature $T_{*}$ for the paramagnon effects we
% study could be much smaller than the Kondo temperature, $T_{K}\sim
% \epsilon_{F}\exp(-1/Jn(\epsilon_{F}))$. 
%In that case, the
% contribution to impurity susceptibility due to paramagnons can be
% much more significant than the usual Kondo contribution to the susceptibility.
% On the other hand if $J<0$, the paramagnon contribution to impurity
% susceptibility still saturates as $T\rightarrow 0$, however the
% impurity susceptibility should not fall below $S^{2}/T$ since the
% ground state of the system has at least a spin $S$. Thus at low
% enough temperatures and for $J<0$, paramagnons might
% not be the most important factor that determine impurity susceptibility. 

%if tunnelling were absent. We believe
%similar results would be obtained for defects with complete rotational symmetry.

We assume that the coupling between a spin and electrons in a single channel
is small, $J n(\epsilon_{F}) \ll 1$. Therefore the terms beyond the Born
approximation can be neglected.\cite{larkin}
Importantly, the overall coupling constant
$g\approx(Jn(\epsilon_{F}))^{2}/(1+F_{a})$ is large in the proximity
of the quantum critical point $F_a \rightarrow -1$.
The magnetic moment of the droplet, $M=(1-J n(\epsilon_F)/(1+F_a))S$, is large,
because the droplet is dressed by (or coupled to) a large number 
of electron channels,  $N_{ch}=1/(1+F_a)$.
Close to the critical point,
the contribution of such magnetic droplets
to the susceptibility and resistivity can  overshadow other
impurity effects.\cite{larkin}

The weak damping $(g\ll 1)$ limit of the same problem was first studied
by Larkin and Melnikov (LM)\cite{larkin}. 
%Large quantum fluctuations
%of the droplet are important here at all temperatures. 
Both resistivity
and susceptibility were found to \emph{decrease} logarithmically with
temperature irrespective of the sign of exchange coupling $J$ as
opposed to the usual Kondo effect where resistivity increases logarithmically
with temperature if $J>0.$
%The immediate purpose of our study is, doubtless, to understand the
%significance of quantum tunnelling for a magnetic impurity with a
%continuous symmetry in a nearly quantum critical metal. 

The rest of this paper is organized as follows. In
Sec.\ref{s:modelformalism} we derive our model for dissipative dynamics
of the magnetic droplet beginning with the exchange Hamiltonian, 
Eq.(\ref{Hex}). The correlation function and susceptibility of the
droplet are studied analytically in Sec.\ref{analysis}. We
study both not-too-low temperatures where tunneling effects are
negligible, and very low temperatures where tunneling makes important 
contributions. Sec.\ref{s:numericalcalculations} provides details of 
the numerical path integral Monte Carlo (PIMC) method we employ. In
Sec.\ref{s:outcome} we present the results of our analytic and numerical
study. Sec.\ref{s:discussion} contains a discussion.

%==============================================================================
\section{Model and Formalism }\label{s:modelformalism}

The magnetic properties of the metal are determined by the spin susceptibility
of the conduction electrons, 
\begin{align*}
D(\mathbf{k},\omega_{n})\delta^{ij} &= 
\int d\mathbf{r}\, e^{i\mathbf{k\cdot\mathbf{r}}}
\int_{0}^{\beta}d\tau\, e^{i\omega_{n}\tau}\langle\sigma^{i}(0,0)
\sigma^{j}(\mathbf{r},\tau)\rangle,
\end{align*}
where $\sigma^{i}(\mathbf{r},\tau)$ 
is the conduction electron spin
density, 
\begin{align*}
\sigma^{i}(\mathbf{r},\tau) &= 
c_{\mu}^{\dagger}(\mathbf{r},\tau)\sigma_{\mu\nu}^{i}
c_{\nu}(\mathbf{r},\tau),
\end{align*}
and $\omega_{n}=2\pi Tn.$ The static part of the susceptibility
is $D(0,0)=2n(\epsilon_{F})/(1+F_{a})$ in terms of the standard Landau
Fermi-liquid parameter $F_{a}.$ Close to a ferromagnetic instability,
$F_{a}\approx-1,$ due to the  large static magnetic susceptibility the impurity
induces a large magnetic droplet with effective moment 
$M=S[1-Jn(\epsilon_{F})/(1+F_{a})].$
In the case of an antiferromagnetic exchange coupling ($J>0$),
the droplet's magnetic moment is polarized in the opposite direction to the
impurity spin, while in the case of ferromagnetic coupling ($J<0$)
the droplet's magnetic moment and the impurity spin are locked parallel
to each other.
The low-lying magnetic excitations of the conduction electrons (which
also constitute the major part of the droplet) are strongly damped,
as can be seen in the expression for spin susceptibility at small
$\mathbf{k}$ and $\omega_{n}$,
\begin{align}
D(\mathbf{k},\omega_{n}) &\approx 
\frac{2n(\epsilon_{F})}{1+F_{a}+(\xi_{0}\mathbf{k})^{2}+
\frac{\pi}{2}\frac{|\omega_{n}|}{\mathbf{k\cdot v}_{F}}};
\label{D}
\end{align}
$\xi_{0}$ is a length scale of the order of interatomic distances.
The dynamics of the impurity are determined by the local susceptibility
$D(\omega_{n})$,\cite{larkin,moriya}
\begin{align}
D(\omega_{n})-D(0) &= \int(d^{3}\mathbf{k})~
\big[ D(\mathbf{k},\omega_{n})-D(\mathbf{k},0) \big]\nonumber \\
 &\approx -\frac{\pi n(\epsilon_{F})^{2}|\omega_{n}|}
{2(\xi_{0}k_{F})^{2}(1+F_{a})}.
\label{Dyn1}
\end{align}
Eq.(\ref{Dyn1}) is valid at low enough frequencies, 
\begin{align*}
|\omega_{n}| & \ll  E_{0}\equiv \epsilon_{F}(1+F_{a})^{3/2}.
\end{align*}
At higher frequencies, $E_{0}\ll|\omega_{n}|\ll\epsilon_{F},$ 
Eq.(\ref{Dyn1}) should be replaced with 
\begin{align}
D(\omega_{n})-D(0) &\approx -\frac{2^{7/3}\pi^{4/3}
\epsilon_{F}n(\epsilon_{F})^{2}}{3^{3/2}(\xi_{0}k_{F})^{8/3}}
\left(\frac{|\omega_{n}|}{\epsilon_{F}}\right)^{1/3}.
\label{Dyn2}
\end{align}

Integrating out the conduction electrons (in the perturbation series of $(Jn(\epsilon_F))^2 \ll 1)$ 
in Eq.(\ref{Hex}) results in a dissipative action for the impurity,\cite{larkin} 
\begin{align*}
S_{imp} [\mathbf{S}] &= \frac{J^{2}}{2}\int_{0}^{\beta}d\tau\, 
d\tau'\mathbf{S}(\tau)\cdot\mathbf{S}(\tau')D(\tau-\tau').
\end{align*}
We are interested in the impurity dynamics at low temperature, so
using Eq.(\ref{Dyn1}) for the interaction, 
\begin{align}
S_{imp} [\mathbf{n}]&= \frac{\pi gT^{2}}{2}\int_{0}^{\beta}d\tau\, 
d\tau'\frac{1-\mathbf{n}(\tau)\cdot\mathbf{n}(\tau')}
{\sin^{2}(\pi T(\tau-\tau'))}.
\label{Sn}
\end{align}
In Eq.(\ref{Sn}), 
$\mathbf{S}(\tau)=S\mathbf{n}(\tau),\,\mathbf{n}(\tau)^{2}=1,$
and we assume that the damping $g$ given by 
\begin{align}
g &= \frac{\pi}{2(1+F_{a})}
[JSn(\epsilon_{F})]^{2},
\label{gdef}
\end{align}
is large $(g>1).$ This is always possible sufficiently close
to the transition. The form of the interaction we chose
in Eq.(\ref{Sn}) is valid only up to an energy $E_{0}.$ We may impose
this cutoff through an additional regularizing term in the action,
\begin{align}
S_{reg} [\mathbf{n}]&= \frac{1}{4E_{0}}\int_{0}^{\beta}d\tau\,
(\partial_{\tau}\mathbf{n})^{2}.
\label{cutoff}
\end{align}
This method of imposing the cutoff is not unique. For instance,
an equally valid option would have been to introduce a 
short time cutoff, $\tau_c$ in
the interaction, $gT^2/\sin^2(\pi T\tau)\rightarrow gT^2/\sin^2(\pi T {\sqrt
  {\tau^2 + \tau_{c}^{2}}})$. The cutoff appears in the correlation
functions only a function of $E_{0}\tau$ or $\tau/\tau_{c}$, and as
long as $gE_{0}\tau \gg 1$, the results are independent of the manner in
which the cutoff is imposed. Parametrizing 
$\mathbf{n}(\tau)=(\cos\varphi(\tau),\sin\varphi(\tau)),$
our dissipative model for the $XY$ impurity takes the final form,
\begin{align}
S[\varphi] &= \frac{1}{4E_{0}}\int_{0}^{\beta}
d\tau\, (\partial_{\tau}\varphi)^{2}+M\int_{0}^{\beta}d\tau\,
\mathbf{h}_{\perp}(\tau)\cdot\mathbf{n}(\tau)+\nonumber \\
 &  +\frac{\pi gT^{2}}{2}\int_{0}^{\beta}d\tau\, 
d\tau'\frac{1-\cos(\varphi(\tau)-\varphi(\tau'))}
{\sin^{2}(\pi T(\tau-\tau'))},
\label{millis1}
\end{align}
where $M=S[1-Jn(\epsilon_{F})/(1+F_{a})]$ is the bare droplet moment,
and $\mathbf{h}_{\perp}$ is an in-plane magnetic field. The Matsubara
fields $\varphi(\tau)$ satisfy periodic boundary conditions up to multiples of $2\pi$: $\varphi(\tau+\beta)=\varphi(\tau) + 2\pi k$, where $k$ is an
integer called the winding number.  The full phase $\varphi(\tau)$ can always be written in the form $\varphi(\tau)=2\pi Tk\tau+\phi(\tau)$, where the residual phase obeys periodic boundary conditions $\phi(\tau+\beta)=\phi(\tau)$.

%as a sum of ``winding number path'' $2\pi T k\tau$ and the residual phase:
 
We are mostly concerned here with the rotational (or orientational)
motion of the droplet's moment.
We assume throughout the main text (but see the concluding section) that
the exchange coupling is sufficiently strong to suppress the fluctuations 
of the amplitude of the bare magnetic moment $M$, namely, 
in the quantum limit or at low temperatures,
$JS \gg kT$ as well as $E_0 \gg kT$.
It also suits us to
regularize the tunneling action through the introduction of a kinetic
term because the model in Eq.(\ref{millis1}) appears in numerous contexts.
We have already mentioned the recent work by Millis and
co-workers\cite{millis1} in which the authors studied the dynamics of
magnetic defects in nearly quantum-critical metals, where the defects
are regions of ordered phase formed due to a local enhancement of the transition
temperature. For defects with $XY$ symmetry, they arrived at
essentially the same strongly-damped model as Eq.(\ref{millis1}), although in their model the kinetic term did not appear simply as a means for imposing a cutoff but had a definite physical meaning as the contribution to the droplet
action from the magnon part of the dispersion curve. Our analysis should be 
valid for such systems as well. 
The action in Eq.(\ref{millis1}) 
%is also identical to the action in Bascones et
%al.\cite{bascones}, which  
also arises from an Ambegaokar-Eckern-Sch\"on\cite{ambegaokar} 
treatment of tunneling through a quantum dot.  
There, the physical meaning of $E_0$ is the charging energy for the 
quantum dot.  

In this paper we study the impurity spin correlator $C(\tau),$
\begin{align}
C(\tau) &= 
\langle\cos(\varphi(\tau)-\varphi(0))\rangle,
\label{ctaudef}
\end{align}
and the zero-frequency impurity susceptibility $\chi_{dc}(T),$
\begin{align}
\chi_{dc}(T) &= M^{2}\int_{0}^{\beta}d\tau~\langle\mathbf{n}
(\tau)\cdot\mathbf{n}(0)\rangle\nonumber \\
 &= M^{2}\int_{0}^{\beta}d\tau~\langle\cos(\varphi
(\tau)-\varphi(0))\rangle.
\label{suscept1}
\end{align}
Calculation of the imaginary part of the susceptibility as well as
transport properties like resistivity involve
subtleties associated with analytic continuation to real
frequencies. These will be studied in a later work. 
In this paper we calculate only the real part of the
susceptibility as shown in Eq.(\ref{suscept1}).

%==============================================================================
\section{Analysis of Impurity spin correlation function and
  susceptibility, $g\gg1$}\label{analysis}

A large value of $g$ tends to suppress large
fluctuations (tunneling) of the droplet moment. Physically, 
the droplet couples to a large number of channels $(g \propto N_{ch})$ of the
conduction-electron continuum, which makes spin flips difficult. We
show below that this is not the case at very low temperatures, where
spin flips (we use the terminology of spin flips and tunneling interchangeably
in order to discuss $XY$ and Ising symmetry simultaneously) can occur even for large $g$. In the first part of this
section, we consider not-too-low
temperatures where tunneling may be disregarded. In the latter half of this
section, we discuss the limits beyond which tunneling may not be
ignored, and analyze the effect of tunneling on the correlation
function and susceptibility. 

%------------------------------------------------------------------------------
\subsection{Ignoring winding numbers}

Let us begin by studying the action, Eq. (\ref{millis1}) (with $\mathbf{h}_{\perp}=0$), by ignoring winding numbers $(k=0)$ and
expanding $\cos(\varphi(\tau)-\varphi(\tau'))$ to quartic
order in the phase difference. We then gather the Gaussian terms
in the resulting action as the bare term and treat the quartic
term as an ``interaction'' $(S=S^{\text{gauss}}+S^{\text{int}}):$
\begin{align}
S^{\text{gauss}}[\phi] &= \frac{1}{4E_{0}}\int_{0}^{\beta}
d\tau(\partial_{\tau}\phi)^{2}+ \nonumber \\ 
 & + \frac{\pi gT^{2}}{4}\int_{0}^{\beta}d\tau\, d\tau'\frac{(\phi(\tau)-\phi(\tau'))^{2}}{\sin^{2}(\pi T(\tau-\tau'))},\\
S^{\text{int}}[\phi] &= -\frac{\pi
  gT^{2}}{48}\int_{0}^{\beta}d\tau\, 
d\tau'\frac{(\phi(\tau)-\phi(\tau'))^{4}}{\sin^{2}(\pi
 T(\tau-\tau'))}. \label{quartic}
\end{align}
 The bare term may be diagonalized by going over to the frequency 
representation,
\begin{align*}
\phi(\tau) &= \sum_{n=1}^{\infty}
(a_{n}\cos\omega_{n}\tau+b_{n}\sin\omega_{n}\tau),\,\mbox{thus,}\\
S^{\text{gauss}} [\phi] &= \frac{1}{2}\sum_{n=1}^{\infty}
(a_{n}^{2}+b_{n}^{2})\left[\frac{T\pi^{2}}
{E_{0}}n^{2}+\pi gn\right].
\end{align*}
%
%Here $a_n$ and $b_n$ are the Fourier cosine and sine coefficients of the residual phase $\phi(\tau)$.  

 In the following discussion, we will need the bare Green function, which is given exactly by the following sum:
 \begin{align}
F_{\text{bare}}(\tau) &= \langle(\phi(\tau)-\phi(0))^{2}\rangle_{\text{bare}}\nonumber \\
 &= 2\sum_{n=1}^{\infty}\frac{1-\cos\omega_{n}\tau}{\frac{T\pi^{2}}{E_{0}}n^{2}+\pi gn}.
\label{e:Fbare}
\end{align}
Essentially, $F_{\text{bare}}(\tau)$ is logarithmic in $\tau$, with a
lower cutoff $\frac{1}{gE_0}$ and an upper cutoff $\frac{1}{2T}$:
	\begin{align}
	F_{\text{bare}}(\tau) 
	&\approx & 
	\begin{cases}
	2E_{0}|\tau|, 
		& \tau\ll\frac{1}{gE_{0}}
		\\
	\frac{2}{\pi g}\ln (2e^\gamma gE_0\tau), 
		& \frac{1}{gE_{0}} \ll \tau \ll \frac{1}{2T}	
		\\
	\frac{2}{\pi g}\ln \frac{2e^{\gamma} gE_{0}\sin \pi T \tau}{\pi T}, 
		& \tau\gg\frac{1}{gE_{0}} 
	\end{cases}
	\end{align}

Consider now the renormalization of $S^{\text{gauss}}$ by $S^{\text{int}}$.
By contracting two of the four fields appearing in $S^{\text{gauss}}$
using the bare Green function, we obtain a one-loop renormalization of the bare
action by the interaction: 
\begin{widetext} 
\begin{align*}
\delta S^{\text{gauss}}_{\text{1 loop}}[\phi] &= -\frac{\pi
  gT^{2}}{8}\int_{0}^{\beta}d\tau\, 
d\tau'\frac{(\phi(\tau)-\phi(\tau'))^{2}}{\sin^{2}(\pi
  T(\tau-\tau'))}\langle(\phi(\tau)-\phi(\tau'))^{2}\rangle_{\text{bare}}\\
 &= -\frac{\pi gT^{2}}{4}\sum_{n=1}^{\infty}\int_{0}^{\beta}d\tau\,d\tau'
\frac{1-\cos(\omega_{n}(\tau-\tau'))}{\frac{T\pi^{2}}{E_{0}}n^{2}+\pi gn}
\times\frac{(\phi(\tau)-\phi(\tau'))^{2}}{\sin^{2}(\pi T(\tau-\tau'))}\\
% &= -\frac{\pi g}{4}\sum_{n=1}^{\infty}\sum_{p=1}^{\infty}(a_{p}^{2}+b_{p}^{2})\frac{1}{\frac{T\pi^{2}}{E_{0}}n^{2}+\pi gn}[2p-|n+p|-|n-p|+2n]\\
% &= -\frac{\pi g}{4}\sum_{p=1}^{\infty}(a_{p}^{2}+b_{p}^{2})\left[\sum_{n=1}^{p}\frac{2n}{\frac{T\pi^{2}}{E_{0}}n^{2}+\pi gn}+\sum_{n=p}^{\infty}\frac{2p}{\frac{T\pi^{2}}{E_{0}}n^{2}+\pi gn}\right]\\
 &\approx
 -\frac{1}{2}\sum_{p=1}^{\infty}
 \left(a_{p}^{2}+b_{p}^{2}\right) p \,\ln \frac{gE_{0}}{Tp} .
\end{align*} 
\end{widetext}
The effective Gaussian action, 
$S^{\text{gauss}}_{\text{eff}} [\phi] = S^{\text{gauss}} [\phi] + \delta
S^{\text{gauss}} [\phi]$, with fluctuations considered up to one loop
is therefore
\begin{align}
S^{\text{gauss}}_{\text{eff}} [\phi] &=
\frac{1}{2}\sum_{n=1}^{\infty} \left(a_{n}^{2}+b_{n}^{2}\right)
\left[\frac{\pi^{2}T}{E_{0}}n^{2}+\pi ng_\text{ren}\right],
\label{Seff}
\end{align}
 where the effective coupling to one loop,
\begin{align}
g_\text{ren}(n) = g\left(1-\frac{1}{\pi g}\ln \frac{gE_{0}}{nT} \right),
\label{Grenorm}
\end{align}
is smaller than the bare coupling $g,$ and the reduction is
strongest at low $n.$ Since $g\propto M$, this means that low-energy paramagnon
fluctuations of the conduction electrons that constitute a large
fraction of the droplet cause a logarithmic reduction of its
moment. 

The phase correlation function with the dressed
coupling $g_\text{ren}(n),$
\begin{align}
C(\tau) 
&= \langle\cos(\phi(\tau)-\phi(0))\rangle=
	\exp(-\langle(\phi(\tau)-\phi(0))^{2}\rangle/2)
	\nonumber \\
 &= \exp\left[-\sum_{n=1}^{\infty}\frac{1-\cos(\omega_{n}\tau)}
	{\frac{\pi T}{E}n^{2}+\pi gn \left(1-\frac{1}{\pi g}\ln \frac{gE_0}{nT} 	\right)}\right]
\nonumber \\
%	 &= \exp\left[\ln\left\{ 1-\frac{1}{\pi g}\ln(gE_{0}|\tau|)\right\} 	\right]
% \nonumber \\
 &= 1-\frac{1}{\pi g}\ln(gE_{0}|\tau|) + O(g^{-3}\ln^{3}(gE_{0}|\tau|)),
\label{Ctau}
\end{align}
 differs at $O(g^{-2})$ from the bare correlator 
\begin{align}
C_{\text{bare}}(\tau) &= \exp(-F_{\text{bare}}(\tau)/2) \sim (gE_{0}|\tau|)^{-\frac{1}{\pi g}}.
\label{Cbare}
\end{align}
 At large $|\tau|,$ the difference between the bare correlator and
dressed correlator can be substantial.  
Eq.(\ref{Ctau}) is a one-loop calculation for $C(\tau)$. Notice
that if the effective $g_{ren}$ has only a logarithmic correction
as in Eq.(\ref{Grenorm}), the correlator $C(\tau)$ has the same 
logarithmic correction as in Eq.(\ref{Ctau}). 
%We argue below that the perturbation theory in $1/g$ does not make any
%contribution to Eq.(\ref{Ctau}) beyond $O(1/g)$. 
%In Eq.(\ref{Seff}) we identify the self energy,
%
%\begin{align*}
%\Sigma(n) & \equiv 
%G_{\mbox{bare}}^{-1}(n)-G^{-1}(n)=n\ln\left(gE_{0}/nT\right),
%\end{align*}
%
% where $G(n)=\langle a_{n}^{2}\rangle=\langle b_{n}^{2}\rangle.$
To obtain the next $O(g^{-1})$ contribution to the coupling $g_{ren}$ in the Gaussian approximation, we need
to calculate the two-loop diagrams.  These include only the second-order diagrams from the
quartic interaction in Eq.(\ref{quartic}), but also one diagram from
the sixth-order fluctuation expansion of the tunneling term in Eq.(\ref{millis1}).
One may check that the most singular contribution in both classes of
diagrams is $O(g^{-1}\ln^2(gE/T))$. What is also clear that
all other diagrams are of the order 
%either $O(g^{-m}\ln^m (gE/T))$ or 
$O(g^{2-m-l}\ln^{m}(gE/T))$, where $m,l$ are natural numbers. 
Therefore if $\ln(gE/T) \ll g$,
Eqs.(\ref{Grenorm},\ref{Ctau}) are good approximations. Since the logarithm
is a slow function of $\tau$ or $T$, there is a large range of $\tau$ or $T$ where $C(\tau)$
can be approximated by Eq.(\ref{Ctau}). 
%There is a simple way to see that such diagrams will cancel.
%By differentiating the tunneling term in Eq.(\ref{millis1}) twice with respect to $(\phi(\tau)-\phi(\tau')),$
%we infer the relation, $g_\text{ren}(\tau)/g=C(\tau).$ Since $C(\tau)$
%has no $O(g^{-2})$ contribution, $g_\text{ren}/g$ doesn't either.
%However if $g_\text{ren}/g$ doesn't have an $O(g^{-2})$
% contribution, the $O(g^{-1})$ contribution to self energy must
% vanish, and moreover, in that case
%one may check that $C(\tau)$ will not have any $O(g^{-3})$ contribution.
%Carrying this sequence forward, we reach the conclusion that if
% winding 
%numbers are absent, then Eq.(\ref{Ctau}) is exact to all orders in
%perturbation theory in $1/g$. 
Such a perturbative analysis does not
preclude a qualitatively different behavior of $C(\tau)$ at long $\tau$'s,
which indeed occurs as we discuss in the next section.
% Our numerical
%calculations show that $C(\tau)$ at large values of $\tau$ 
%starts significantly deviating from Eq.(\ref{Ctau}) when $C(\tau)$ drops below $1/2$.
%We find (see below) it is described well by
%$C(\tau)\approx (T/T_{*})^{2}/\sin^{2}(\pi T\tau),\, T_{*}\sim
%E_{0}e^{-\pi g/2}$.  
Eq.(\ref{Ctau}) leads to a magnetic susceptibility, 
\begin{align}
\chi_{dc}(T) 
&\approx \frac{M^{2}}{T}\left[1-\frac{1}{\pi g}\ln \frac{gE_0}{T}\right].
\label{chiEfetov}
\end{align}
In deriving Eq.(\ref{chiEfetov}), we assumed that $C(\tau)$ is
effectively constant over the whole of the interval $[0,\beta]$, and
that the errors in this assumption simply modify the cutoff of the logarithm.

At very large $|\tau|\gg 1/T_{gauss} \sim (gE_{0})^{-1}e^{\pi g},$ $C(\tau)$ in Eq.(\ref{Ctau})
becomes unphysically negative, which sets a low-temperature limit of
validity of our perturbative analysis. 
Similarly, Eq.(\ref{chiEfetov}) is unable to resolve the question as to how the susceptibility
behaves as $T\rightarrow0$ since our present perturbation theory, which
ignores winding numbers and higher-order residual fluctuations,
breaks down at temperatures $T\lesssim T_{gauss}=gE_{0}e^{-\pi g}.$
Near $T=T_{gauss}$, Eq.(\ref{chiEfetov}) gives an exponentially large susceptibility,
\begin{align*}
\chi_{dc}(T_{gauss}) &= \frac{M^{2}(e^{\pi g}-1-\pi g)}{\pi
  g^{2}E_{0}}.
\end{align*} 
At lower temperatures $T<T_{gauss}$ one must consider winding numbers,
higher-order residual fluctuations, and non-perturbative (in $1/g$) contributions
to $\chi_{dc}$. Our numerical calculations show that for $g\gtrsim1,$
$C(\tau)$ visibly falls below Eq.(\ref{Ctau}) around $C(\tau)\approx1/2,$
so Eq.(\ref{Ctau}) becomes inaccurate even before $|\tau|\approx 1/T_{gauss}$
is reached.
 
We emphasize that the logarithmic temperature dependence
in Eq.(\ref{chiEfetov}) is not an approximate expansion of $\frac{M^{2}}{T}(T/gE_{0})^{1/(\pi g)}$
to order $1/g$, but is in fact more accurate than this, in the range $gE_{0}\gg T\gg T_{gauss}$.

%------------------------------------------------------------------------------
\subsection{Taking winding numbers into account}

We now turn our attention to paths $\varphi(\tau)=\omega_k\tau+\phi(\tau)$ with a finite winding number $k$.  We have chosen to expand the full phase $\varphi$ in terms of residual-phase fluctuations, $\phi$, about the `classical' paths $\varphi^\text{cl} (\tau) = \omega_k \tau$.  There are reasons for doing so.  Firstly, these classical paths are solutions of the Euler-Lagrange equations, so they are \emph{stationary points} of the action $S_k[\phi]$.  Secondly, expanding the action Eq.\eqref{millis1} to second order in the fluctuations shows that all the `spring constants' (the coefficients of ${a_n}^2$ and ${b_n}^2$) are positive, so the classical paths are \emph{local minima} of the action:
\begin{widetext} 
\begin{align}
S^{\text{gauss}}_k [\phi] &= \frac{\pi^{2}T}{E_{0}}k^{2}+\pi g|k|+
\frac{1}{4E_{0}}\int_{0}^{\beta}d\tau\,(\partial_{\tau}\phi)^{2}+
%\nonumber \\& +
\frac{\pi g}{4}\int_{0}^{\beta}d\tau\, d\tau'
\frac{\cos(2\pi kT(\tau-\tau'))}{\sin^{2}(\pi T(\tau-\tau'))}(\phi(\tau)-\phi(\tau'))^{2}
\nonumber \\&= 
 \frac{\pi^{2}T}{E_{0}}k^{2}+\pi g|k|+
 %\nonumber \\ &+
\sum_{n=1}^{\infty}
\frac{1}{2}
\left[\frac{\pi^{2}T}{E_{0}}n^{2}+\frac{\pi
 g}{2}\left(|n+k|+|n-k|-2|k|\right)\right]
 (a_{n}^{2}+b_{n}^{2})
 .
\label{Swinding}
\end{align}
\end{widetext} 
Thirdly, we believe (although we have not proved) that for each value of $k$, 
the classical paths $\varphi^\text{cl}_k (\tau)$ is \emph{the unique global minimum} of $S_k[\phi]$.

The Fourier modes form a complete basis, so our parametrization
automatically encompasses the Korshunov instanton trajectories used by
other authors.\cite{korshunov}  Indeed, it has been found that instanton techniques, 
while useful in Josephson-junction problems, have limited applicability in the current situation 
because of the failure of the `non-interacting instanton gas approximation'.\cite{falci95}

At first sight the presence of $\pi g|k|$ in Eq.(\ref{Swinding})
seems to suggest that a large value of $g$ rules out significant finite-winding-number effects. 
Observe however that the fluctuation term also depends
on $k,$ so we should integrate out the residual-phase fluctuations
and find whether the resulting $k$-dependent contribution to the
action encourages or further discourages finite winding numbers. 
In Eq.(\ref{Swinding}), the contribution to residual fluctuations
arising from the tunneling term vanishes for Matsubara frequencies $n \leq
|k|$. As a result, cubic and higher order residual phase
fluctuations begin playing a role.
However despite our considering residual fluctuations 
only to Gaussian order, we find that
at not too low temperatures, 
exact numerical calculations (discussed in
Sec.\ref{s:numericalcalculations}) are in good
agreement with our result in Eq.(\ref{e:Swinding2}) and Eq.(\ref{gamma})
below. For very low temperatures, the Gaussian expansion is not accurate. 

Let us now integrate out the residual phases, $a_n$ and $b_n$, from Eq.(\ref{Swinding}), to obtain the `effective winding-number action' $S^\text{gauss}_k$ in the Gaussian approximation.  The functional integration produces a determinant
\[ D_k = \prod_{n=1}^{\infty}\left[\frac{T\pi^{2}}{E_{0}}n^{2}+
\frac{\pi g}{2}(|n+k|+|n-k|-2|k|)\right]^{-1} .\]
To eliminate the divergences that are inherent in the definition of the path integral, we normalize this against the determinant $D_0$ corresponding
to zero winding number.  The fluctuation contribution to the effective winding-number action is thus

\begin{widetext} 

	\begin{align}
	\Delta S^\text{gauss}_k 
	&= S^\text{gauss}_k - S^\text{gauss}_0 = \ln D_k/D_0 \\
	&=\frac{\pi^{2}T}{E_{0}}k^{2}+\pi g|k|
	+	\sum_{n=1}^{\infty}
		\frac{1}{2}	\ln
		\frac{
			\frac{\pi^{2}T}{E_{0}}n^{2}+\frac{\pi
			 g}{2}\left(|n+k|+|n-k|-2|k|\right)
		}{
			\frac{\pi^{2}T}{E_{0}}n^{2}+\pi g|n|
		}
		\\	 
	&=\frac{\pi^{2}T}{E_{0}}k^{2}+\pi g|k|
		-	\ln \frac{ 
				\Gamma\left(1+|k|+\frac{x-\sqrt{x}\sqrt{x+4|k|}}{2}\right) ~
				\Gamma\left(1+|k|+\frac{x+\sqrt{x}\sqrt{x+4|k|}}{2}\right)
			}{\Gamma(1+|k|) ^2~ \Gamma(1+x)}
	\label{e:Swinding2}
	\end{align}

\end{widetext} 
where $x=\frac{g}{2\pi m T}$.  If $g$ is large, Stirling's approximation leads to
	\begin{align*}
	\Delta S^\text{gauss}_k 
	&\approx \frac{\pi^{2}T}{E_{0}}k^{2} + \pi \gamma^\text{gauss}_k |k|
	\end{align*}	
where $\gamma^{\text{gauss}}$ is a temperature- and winding-number-dependent effective coupling parameter given by 
	\begin{align}
	\gamma^{\text{gauss}}_k 
	&= 
		g\left(1-\frac{2}{\pi
    g}\ln \frac{gE_{0}}{T|k|} \right)
	 + \text{const}
	 ,
	\label{gamma}
	\end{align}
where the `constant' is in $O(g^0)$, but may depend on $k$.
%where the superscript on $\gamma^{\text{gauss}}$ reminds that $\gamma$ was calculated considering Gaussian fluctuations around winding number trajectories.
 Observe that $\gamma^{\text{gauss}}$ has a form very similar to the fluctuation-dressed coupling constant $g_\text{ren}(n),$ 
\[
g_\text{ren}(n) = g\left(1-\frac{1}{\pi g}\ln \frac{gE_{0}}{nT} \right),
\]
 obtained by disregarding winding numbers, and also that to leading
order, $\gamma^{\text{gauss}}<g_\text{ren}.$ We note in Eq.(\ref{e:Swinding2})
that when $\pi\gamma=1,$ the phase correlation function evaluated
ignoring winding numbers, 
$C(\tau)\approx g_\text{ren}/g=\frac{1}{2}-\frac{1}{2\pi g}$
is approximately $1/2.$ Therefore once $|\tau|$ is large enough
such that $C(\tau)\lesssim1/2,$ Eq.(\ref{Ctau}) is no longer accurate,
and the effect of winding numbers \emph{must} be considered even if
$g\gg1.$   The criterion $\pi\gamma=1$ sets a crossover temperature
$T_{*} \sim gE_0 e^{-\pi g/2}$ marking the onset of tunneling
effects. This is consistent with numerical calculations that indicate
\begin{align}
T_{*} &= E_0/ \sqrt{2 (e^{\pi g} - 1)} .  % gE_{0}e^{(\pi g-1)/2}.
\label{Tstar}
\end{align}

For calculating correlation functions at low temperature, winding numbers
as well as higher order residual fluctuations need
to be considered. 
In the regime $\pi \gamma^\text{gauss} > 1$, the contributions from
winding number trajectories to the correlation function are exponentially
small, and the correlation function behaves according to Eq.(\ref{Ctau}).
In the regime $\pi \gamma^\text{gauss} < 1$ or below and close to $T_{*}$,
both even and odd residual fluctuations are 
important in the action when the winding number is non-zero. 
Due to these nonlinearities, we have not been able to calculate the
correlation function directly beyond
quadratic order. However we have numerical results (see below) for contributions
from residual fluctuations around the winding number trajectories.

Finally, we note here that we have studied the model in
Eq.(\ref{millis1}) from another direction, namely, by doing perturbation theory in powers of $g$, 
as opposed to powers of $1/g$ as in this paper.  The effective action and correlation functions 
thus obtained will be plotted in the results section of this paper for comparison, but the details 
will be described in another paper, since they pertain mainly to the small-$g$ limit which 
is not the focus of this paper.

%==============================================================================
\section{Numerics}\label{s:numericalcalculations}

We now describe our numerical methods.  Since the phase
$\varphi(\tau)$ is a real scalar field  and the action $S[\varphi]$ is
a real functional, the system can be studied using path integral Monte
Carlo simulation (see, for instance, Refs.~\onlinecite{bascones,herrero}).

We first discuss the discretization, which is an inevitable part of 
Monte Carlo simulation. 
The path is sampled at $N$ values of imaginary time, 
$\tau_{j}=j\varepsilon$, where $j=0,1,2,\dotsc,N-1$.  
The kinetic term is discretized using the `primitive approximation',
\cite{ceperley95} that is, 
by assuming that the path interpolates linearly between adjacent sample points.  
This is not as crude as it sounds: for the free quantum rotor, 
the primitive action coincides 
with the exact renormalized action obtained by integrating out all intermediate phases $\varphi_\tau$ 
for $\tau\neq j\varepsilon$. The dissipative term is likewise approximated by a quadrature formula based 
on bilinear interpolation.  The double pole in the dissipation kernel, $\alpha(\tau)=\frac{T^{2}}{\sin^{2}\pi T\tau}$, 
causes the integrand at `diagonal' grid points $i=j$ to be indeterminate.  To deal with this, we transfer 
the quadrature weight symmetrically off the point $i=j$ onto the points $i=j\pm 1$.  
This is equivalent to the scheme of Ref.~\onlinecite{bascones}:
	\begin{align}
	S (\{\varphi\})
	&=
		\tfrac{1}{4E_0\varepsilon} 
		\sum_{i} \left( \varphi_{i} - \varphi_{i+1} \right)^2
		\nonumber\\&{}
		+\pi g \varepsilon^2   
		\sum_{i\neq j} \alpha_{i-j} \sin^2	\tfrac{\varphi_i - \varphi_j}{2} ,
	\label{e:discrete_action}
	\end{align}
where
	\begin{align}
	\alpha_j
	&=\begin{cases}
%		j=0:									\quad&	0 \text{~(irrelevant)} \\
		j=1 \text{~or~} N-1:	\quad&	\tfrac{3}{2} \alpha(j\varepsilon) \\
		j=2,\dotsc,N-2:			\quad&	\alpha(j\varepsilon) .
		\end{cases}	
	\end{align}
Actually the criterion for a good discretization of the action is not 
how well it approximates the action, but how well it reproduces the 
correlation functions.  It is possible to derive a discrete action 
that, when used in PIMC, will produce a correlation function $C_j$
which is \emph{exactly} equal to $C(\tau_j)$ 
\emph{in the Gaussian approximation}.  However, away from the 
Gaussian limit, this approach did not give a significant improvement 
over the others.

 The time-step, $\varepsilon=\beta/N$, is restricted by the smallest
 time scale in the problem, which is the lower cutoff of the 
logarithm in $F_{\mbox{bare}}(\tau)$,
that is, $1/gE_0$.  Ideally $\varepsilon$ should be chosen to be a 
constant multiple of $1/gE_0$, but the thermodynamic integration
 method described later requires the same $\varepsilon$ to be used 
for all values of $g$.  Hence, we have used a time-step 
$\varepsilon \leq 1/8E_0$.  The discretization error in $C(\beta/2)$ 
can be estimated by comparing runs with different values of
 $\varepsilon$, and can be 1\% or more at the largest values of $g$.  
This error, however, is simply a tendency to globally overestimate 
or underestimate $C(\tau)$, corresponding to a small renormalization 
of the parameters $E_0$ and $g$, and does not affect the conclusions 
of this work regarding the asymptotic behavior of $C(\tau)$.

To generate the Markov chain, we use the Hamiltonian Monte Carlo (HMC) 
method.\cite{neal93,mackaybook}  This is a version of the Metropolis 
algorithm in which new configurations are proposed by evolving old 
configurations in phase space according to Hamiltonian dynamics, 
with the aid of fictitious momenta.  The bias introduced during 
inexact Verlet time evolution is compensated exactly by the Metropolis 
rejection step; the scheme can be shown to satisfy detailed balance.  
During each Verlet trajectory, the configuration evolves 
ballistically rather than diffusively as in the standard Metropolis 
method, which allows a much faster exploration of configuration 
space.\cite{ceperley95}  The HMC method is particularly suitable for 
smooth actions such as Eq.\ \eqref{e:discrete_action}, where there are 
no constraints or collisions to complicate the time evolution.  
Although the long-ranged term in the action contains a double sum, 
this can be treated using FFT methods, so that the computational cost 
of each HMC step is $O(N\log N)$ rather than $O(N^2)$.

The HMC method in its basic form still suffers from the problem that
the Verlet time-step must not exceed the period of oscillation of the
short-wavelength Fourier components of the path, and thus the long-wavelength
components take many time-steps to go through one oscillation. The
solution to this is, preconditioning.  One can exploit the freedom in the definition of the fictitious momenta: instead of giving all beads
the same fictitious mass, the Fourier modes are given wavelength-dependent
masses so that all wavelengths oscillate at the same rate.
If the action had been quadratic in $\phi$, then the Fourier components
would represent independent normal modes, and one could achieve complete
randomization in a single trajectory.

The multilevel Metropolis algorithm\cite{ceperley95} 
is a strong contender, and has the advantage that winding-number
changes can be included naturally in the proposal distribution.  
However, at the time of writing we do not know an accurate `level
action', so to achieve acceptable acceptance ratios, the multilevel 
method would have to be applied to short sections of path, 
sacrificing the efficiency of the FFT method.

The algorithm automatically makes jumps between different subspaces.  
However, in order to do so it must pass over an energy barrier between 
the $k=0$ and $k=1$ subspaces proportional to $1/\varepsilon$, 
corresponding to a phase difference of $\pi$ between adjacent
time-slices. This is $N$ times higher than the typical energy 
difference between a $k=1$ path and a $k=0$ path.  Hence,
winding-number changes can be thermodynamically possible but 
kinetically hindered, which is an obstacle to achieving ergodicity.  
One approach we tried is the `clipped-barrier' scheme, which operates 
as follows. During the calculation of the MD trajectory, use a
modified fictitious force, corresponding to an action in which the 
barriers have been clipped.  This makes it easier for the time
evolution to climb over the barriers.  Then, use the true action in 
the Metropolis accept/reject step, which restores detailed balance.  
We found that this method produced shorter autocorrelation times.

Nevertheless, a more attractive approach, which eliminates the above 
problem entirely, is to perform simulations at fixed winding number, 
in separate $k$-subspaces.  We compute the correlation functions at 
fixed winding number, $C_k(\tau)$, for each $k$:
	\begin{align*}
	C_k(\tau)
	&=\frac{1}{Z_k} \int[d\phi]~ e^{-S_k[\phi]}
		\cos(\varphi_\tau - \varphi_{\tau'})  ,
	\end{align*}	
where the partition function within each subspace is
	\begin{align*}
	Z_k = e^{-S_k }
	&= \int[d\varphi]~ e^{-S_k[\phi]}.
	\end{align*}	
The full correlator is given by a weighted average
	\begin{align}
	C(\tau)
	&=\frac{\displaystyle\sum_k e^{-S_k} C_k(\tau)}
		{\displaystyle\sum_k e^{-S_k}}	
	=\frac{\displaystyle\sum_k e^{-\Delta S_k} C_k(\tau)}
		{\displaystyle\sum_k e^{-\Delta S_k}}	
	\label{e:ctau_weighted_average}
	\end{align}
where $\Delta S_k=S_k - S_0$ is the relative effective winding-number action or `free energy difference'.  This can be computed by simple importance sampling,\cite{neal93,herrero} or by thermodynamic integration.\cite{neal93}  We use the latter method, as it is more reliable when $\Delta S_k$ is large.
Define the thermodynamic function $\Psi_k=-\frac{\partial
  S_k}{\partial g}$, 
which can be obtained from $C_k(\tau)$:
	\begin{align*}
	\Psi_k
	&=\frac{\pi\beta}{2} \int_0^\beta d\tau~ \alpha(\tau) 
\left(1 - C_k(\tau) \right).
	\end{align*}	
$\Psi_k$ actually diverges as $\ln N$, where $N$ is the Trotter number
(number of time-slices), but $\Delta\Psi_k = \Psi_k - \Psi_0$ is finite for
large $N$.  The error caused by subtracting two large numbers is not
too serious in practice.  We can now obtain $\Delta S_k$ by integration:
	\begin{align*}
	\Delta S_k(g)
	&=\Delta S_k(0) - \int_0^g dg'~ \Delta \Psi_k(g') .
	\end{align*}	
The effort of computing $C_k(\tau)$ for many intermediate values of
$g$ is well rewarded, because one can then compute $\Psi_k$, $\Delta
S_k$ and $C(\tau)$ for \emph{each} of these $g$'s with no extra work.  
Besides, $\Delta S_k(g)$ is the \emph{true} effective winding-number
action, which can be compared directly with the analytic estimate 
$\Delta S^\text{gauss}_k(g)$, thus providing a valuable check of the 
numerics, and an indication of the range of validity of the analytics.

All the data presented in the next section were computed using the 
fixed-winding-number method.  The computation took the equivalent of 
38 Pentium4 2.0GHz CPU-days.  1200 runs were performed for 12 values 
of $\beta$ from 2 to 768, 10 values of $g$ from $0.125$ to $8$, and 
10 values of $k$ from 0 to 9.  During each run, the system was
equilibrated for 100 HMC steps and the correlation functions were
subsequently averaged over 250,000 to 1,000,000 steps.  
The Verlet time-step was taken to be $0.1$--$0.2$ of the maximum 
allowable time-step, and the average length of each HMC trajectory was 
$10$--$15$ time-steps.   The Metropolis acceptance ratios were 
typically 65\%--95\%.  The autocorrelation time was estimated by 
calculating the autocorrelation of the deviation of $C(\beta/2)$ 
from its mean value, and was of the order of $2$--$10$ HMC steps.  

The Monte Carlo error can be accounted for as follows.  
When $\beta$ is large (768, say) and $g$ is not too large ($0.5$,
say), the phases $\varphi(0)$ and $\varphi(\beta/2)$ are 
practically uncorrelated, so $\cos (\varphi(0)-\varphi(\beta/2))$ 
is almost symmetrically distributed on $[-1,1]$, with a variance of 
the order of 1.  Assume that $C(\tau)$ falls to a negligible value at 
$\tau \sim 4$, so that the interval $[0,768]$ consists of about 200 
independent blocks.  Then, translational averaging reduces the 
variance of the $C(\beta/2)$ estimator for a single configuration 
to about $1/\sqrt{200} \approx 0.07$.  A run of 1,000,000 HMC steps 
with an autocorrelation time of 8 steps is equivalent to 125,000 
independent samples.   Averaging over these reduces the variance 
to $0.07/\sqrt{125000} \approx 0.0002$.  Indeed, the data in 
Figs.~\ref{f:ctau_b0768_gall} and \ref{f:cbot} exhibit MC noise 
of amplitude $e^{-8.5} \approx 0.0002$.

%==============================================================================
\section{Results}\label{s:outcome}

\begin{figure}
\includegraphics{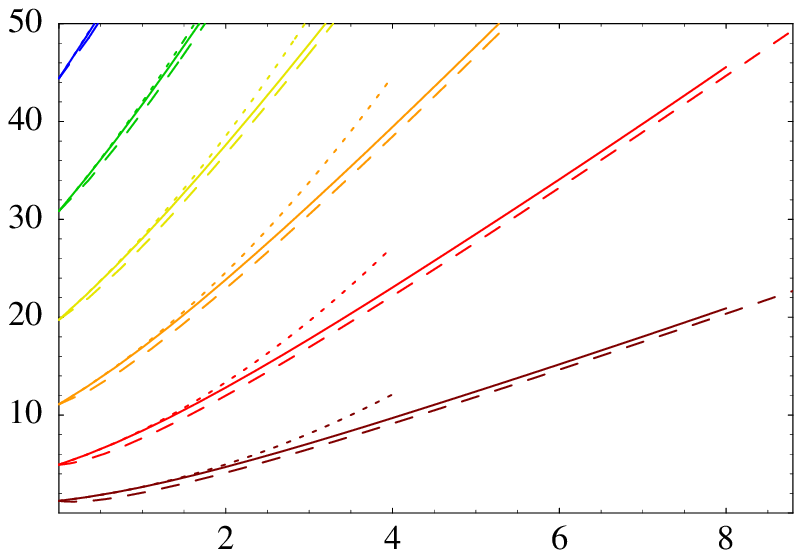}
\includegraphics{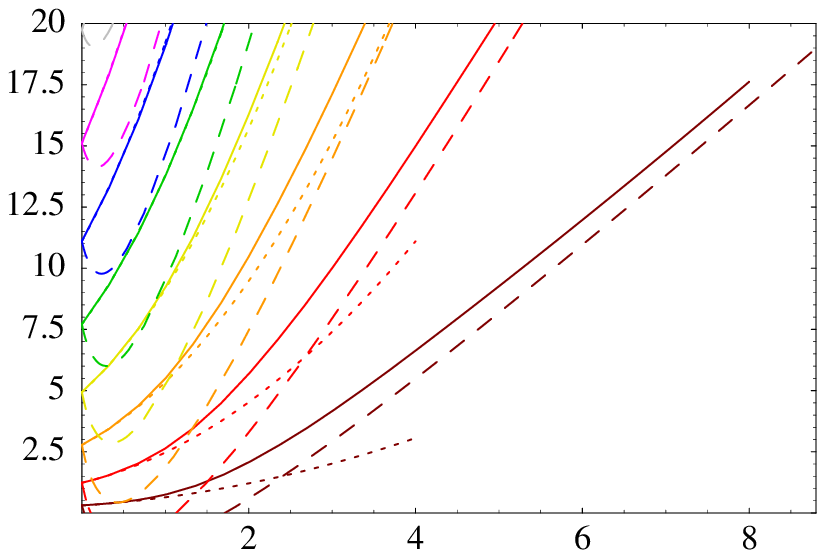}
\includegraphics{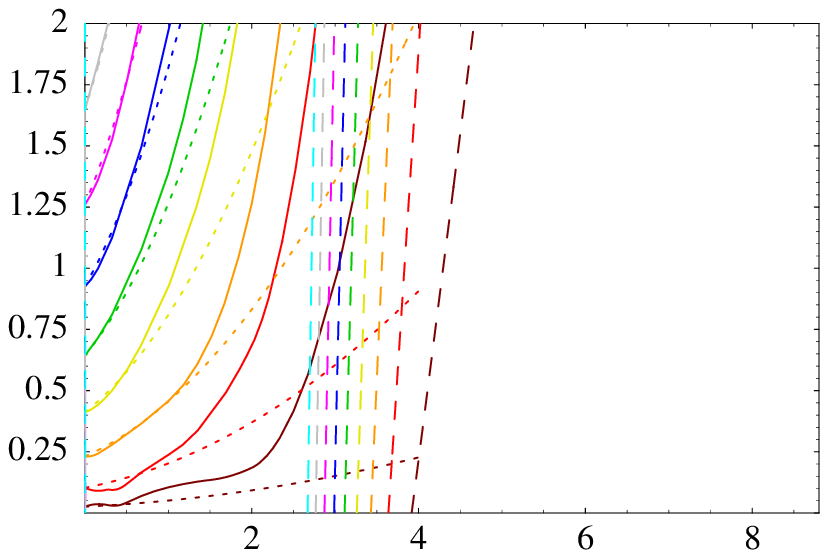}
\caption{\label{f:skg_betaall}
$\Delta S$ vs $g$ for $\beta=16, 64, 768$ (top to bottom).
Resistor color codes indicate $k=1,2,\dotsc,9$.
Solid curves: $\Delta S$ obtained by integration of $\Psi_k$ calculated from PIMC.
Dashed curves: $\Delta S$ evaluated considering only Gaussian
fluctuations about winding number trajectories.
Dotted curves: $\Delta S$ from perturbation theory about small-$g$ limit, up to $O(g^2)$.
At high temperatures (top), PIMC results agree with the Gaussian
approximation in Eq.(\ref{e:Swinding2}). As temperature is lowered (middle),
non-Gaussian fluctuations become important, and significant deviations
from the Gaussian approximation are observed. At very low
temperatures (bottom), Eq.(\ref{e:Swinding2}) fails completely.}
\end{figure}

Figure \ref{f:skg_betaall} shows the relative effective winding-number action, $\Delta S_{k}(g)$.  The numerical results are plotted together with the Gaussian approximation, Eq.\eqref{e:Swinding2}, and the result from small-$g$ perturbation theory.  There is evidently a crossover from one regime to the other.  It is seen that the error in Eq.\eqref{e:Swinding2} is indeed of the form $O(g^0)\times \text{function of $k$}$.

Figure \ref{f:ctau_b0768_gall} shows the correlator $C(\tau)$ vs $\tau$.  
It is possible to identify at least three distinct regimes in the behavior of $C(\tau)$:

\begin{figure*}[p]
\includegraphics{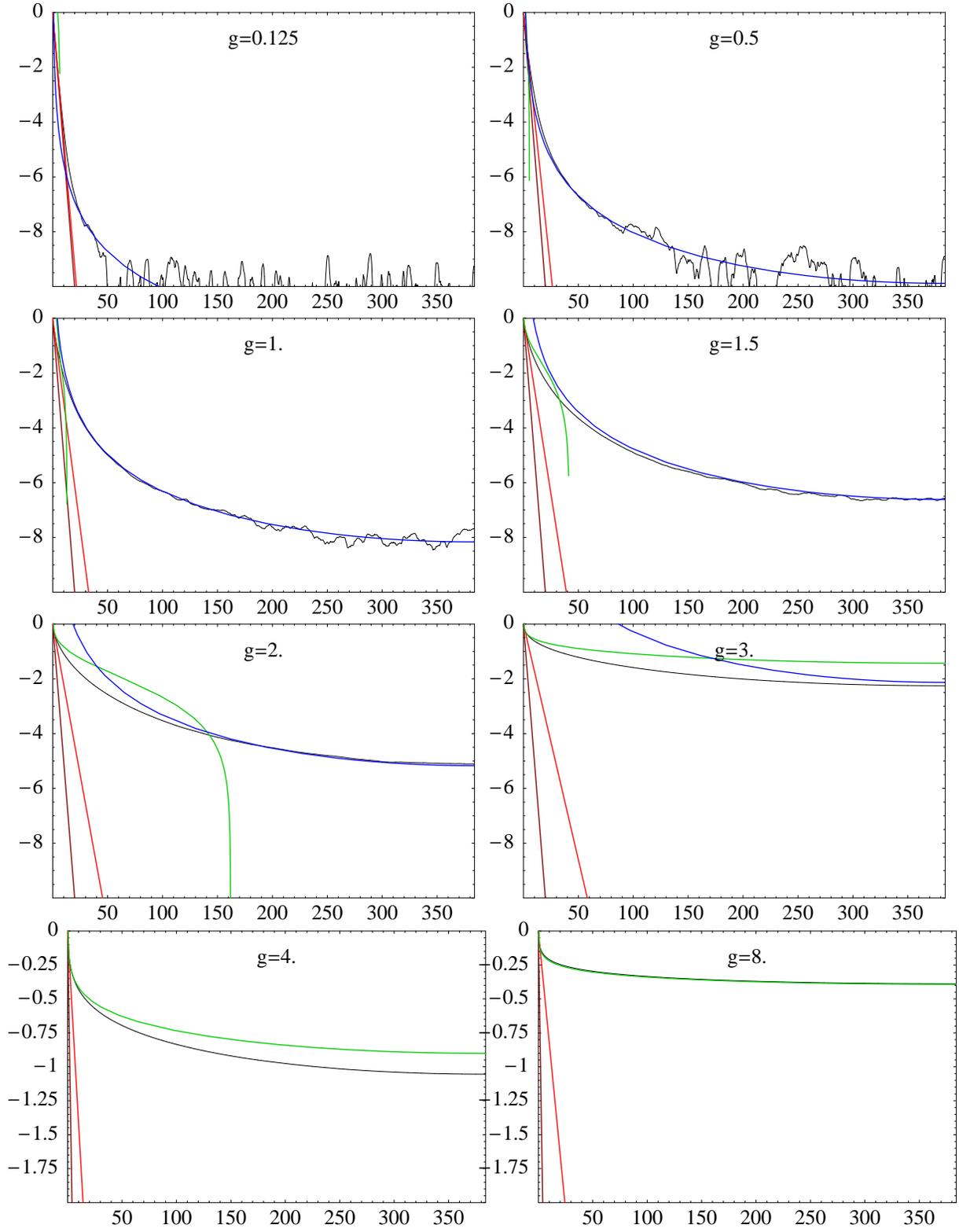}
\caption{\label{f:ctau_b0768_gall} 
$\ln[C(\tau)]$ vs $\tau$ for $\beta=768$ for various values of $g$.  The curves are symmetrical about $\beta/2$.  Black curves: PIMC data.
Brown: $C^\text{exp}$.  Red: $C^\text{renexp}$.  Green:
$C^\text{log}$.  Blue: $C^\text{alg}$.  Below $e^{-8}$ the data is
swamped by Monte Carlo error. For $g\geq 1$, $C(\beta/2)$ evolves
logarithmically (green curve) while $\tau T_{*} <
1,\,T_{*}=E_{0}/\sqrt{2(e^{\pi g} - 1)}$. For $\tau T_{*}\gg 1$, $C(\tau)$
obeys a power-law (blue curve), $C(\tau)= (T/T_{*})^{2}/\sin^{2}(\pi
T\tau)$, for any finite value of $g$.  
}
\end{figure*}

%------------------------------------------------------------------------------
\paragraph{Exponential regime:}

For small $g$ and small $\tau$, the `kinetic' term, Eq.\eqref{cutoff}, dominates, and the system behaves like a free quantum rotor.  The correlation function decays exponentially (brown lines in Fig.~\ref{f:ctau_b0768_gall}):
	\begin{equation}
	C^\text{exp}(\tau) \approx \exp \left(-E_{0}\tau\right).
	\label{e:ctau_exp}
	\end{equation}
For slightly larger $\tau$, theory and numerics suggest exponential decay with a reduced effective $E_0$ (red lines):
	\begin{equation}
	C^\text{renexp}(\tau) \approx \exp \left( -\frac{E_{0}\tau}{1+2g/\pi} \right).
	\label{e:ctau_renexp}
	\end{equation}
The exponential regime is not very relevant in the context of this paper, which is concerned with large $g$.

%------------------------------------------------------------------------------
\paragraph{Logarithmic regime:}

As explained in Eq.(\ref{Ctau}), $C(\tau)$ is logarithmic in the exponentially wide range 
$\frac{1}{gE_{0}}\ll\tau\ll\frac{1}{gE_{0}}e^{\pi g/2}$.  Restoring the correct cutoff in the logarithm,
	\begin{align}
	C^\text{log} (\tau)
	&=1 - \frac{1}{\pi g} \ln \frac{2e^\gamma gE_{0} \sin \pi T
	\tau}{\pi T} ,
	\label{e:ctau_log}
	\end{align}
%and hence in the exponentially wide temperature
%region $\frac{1}{gE_{0}}\ll\beta\ll\frac{1}{gE_{0}}e^{\pi g/2}$
%$C(\beta/2)$ behaves logarithmically: 
%	\begin{equation}
%	C(\beta/2) \approx 1-\frac{1}{\pi g}\ln\frac{2e^{\gamma}g\beta
%	  E_{0}}{\pi}.
%	\label{logregime}
%	\end{equation}
%This implies that $\chi$ behaves logarithmically as well:

%------------------------------------------------------------------------------
\paragraph{Algebraic regime:}

For $\tau \gtrsim \frac{e^{\pi g}}{2e^\gamma gE_0}$, Eq.(\ref{e:ctau_log}) becomes negative, indicating complete failure of the approximation.
In fact, the numerically-obtained $C(\tau)$ begins \emph{falling} below $C^\text{log} (\tau)$ once $C(\tau) \lesssim 1/2$, as nonzero winding-number trajectories start becoming important.  However, this downward trend does not continue indefinitely, but is arrested by an \emph{algebraic} decay.  It turns out that for large $\tau$, the PIMC calculations agree very well with the following prediction of small-$g$ perturbation theory:
	\begin{align}
	C^\text{alg} (\tau)
	&=\rho(g) \frac{(T/2E_0)^2}{\sin^2 \pi T \tau},
	\label{e:ctau_alg}
	\end{align}
where the function $\rho(g)$ is equal to
	\begin{align}
	\rho(g)
	&=8(e^{\pi g} - 1) 
	=8\pi g + 4\pi^2 g^2 + O(g^3).
	\label{e:rho_g}
	\end{align}
In fact, in this regime, the individual $C_k(\tau)$ themselves are also given by Eq.(\ref{e:ctau_alg}).  Thus, although nonzero winding-number trajectories occur in thermodynamic averages with a significant probability, a calculation which ignores them will fortuitously give the right answer.

The $g=2$ and $g=3$ graphs in Fig.~\ref{f:ctau_b0768_gall} clearly show the crossover from the logarithmic regime to the algebraic regime.

Our result in Eq.(\ref{e:ctau_alg}) supports previous 
findings\cite{chakravarty} that the long time dynamics
$(C(\tau)= (T/T_{*})^{2}/\sin^2(\pi T\tau))$ of the Coulomb gas
model, such as Eq.(\ref{millis1}), is determined by the 
critical nature of the Ohmic dissipation. This is also confirmed in
conformal field theory calculations of the spin correlation function
in the Kondo problem.\cite{ludwig} Further
insight comes from Griffiths' theorem\cite{griffiths2} that at large $\tau$, the correlation function $C(\tau)$ cannot decay faster than the interaction. 

Considering a logarithmic law, 
\[C(\tau)=1-(1/\pi g)\ln(gE_{0}\tau),
\]
  for $\tau T_{*} < 1$, and a power-law behavior,
\[C(\tau)=(T/T_{*})^{2}/\sin^{2}(\pi T\tau),
\]
 for $\tau T_{*}>1$, we
  obtain the impurity susceptibility,  
\begin{equation}
\chi_{dc}(T)\approx \chi_{dc}(0) - (2 M^{2}/3T_{*})(T/T_{*})^2,\qquad T \ll T_{*}.
\label{e:chivikram}
\end{equation}
Here $\chi_{dc}(0)$ is exponentially large in $g$, of the order of
$\chi_{dc}(T_{*})$ evaluated using Eq.(\ref{chiEfetov}). 

\begin{figure}
\includegraphics{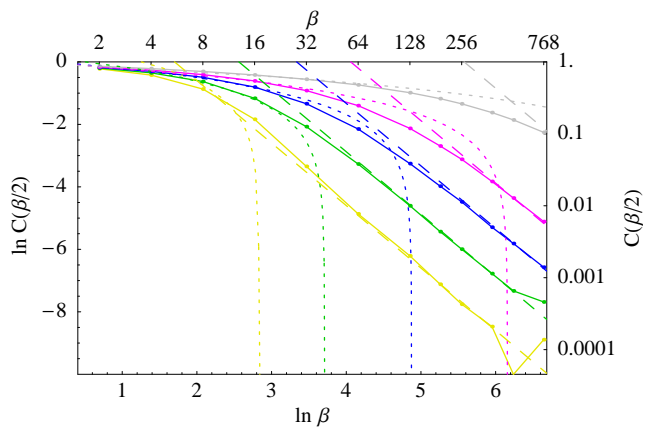}
\caption{\label{f:cbot} 
$C(\beta/2)$ vs $\beta$ for $g=0.5, 1, 1.5, 2, 3$ (bottom to top).
  Colors indicate values of $g$.  Solid lines are PIMC results.
  Dotted lines are $C^\text{log}(\beta/2)$.  Dashed lines are
  $C^\text{alg}(\beta/2)$. A clear crossover from a
  logarithmic law to a power law can be seen as the temperature is decreased.  
}
\end{figure}

We also present a few more figures.  Figure \ref{f:cbot} is a log-log plot of $C(\beta/2)$ vs $\beta$ for various $g$, clearly showing the crossover from the logarithmic regime to the algebraic regime.

\begin{figure}
\includegraphics{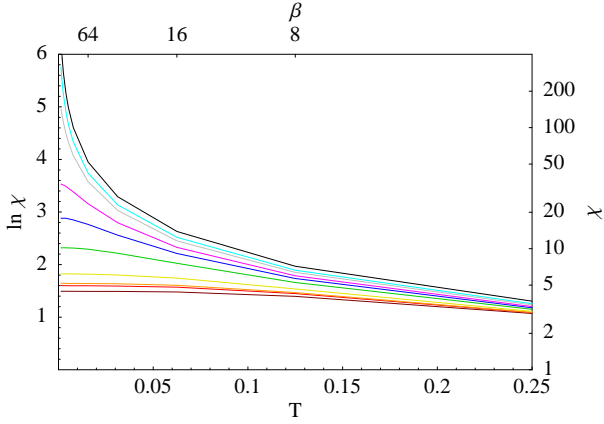}
\caption{\label{f:chi} $\chi_{dc}$ vs $T$ for $g=0.125, 0.25, 0.3, 0.5, 1, 1.5, 2, 3, 4, 8$ (bottom to top). 
The saturation of impurity susceptibility below $T^*$ and a $T^{2}$ deviation from saturation can be seen in all but the top three curves, for which $T^*$ is extremely small.}
\end{figure}

Figure \ref{f:chi} is a log-log plot of $\chi(T)$ vs $\beta$ for various $g$, confirming that at exponentially low temperatures the susceptibility saturates at exponentially large values according to Eq. \eqref{e:chivikram}.  Note that for the larger values of $g$, simulations have not been performed at low enough $T$ to observe saturation.  The errors in $\chi(T)$ are smaller than the errors in $C(\beta/2)$ because $\chi$, being an integral, is dominated by the behavior of $C(\tau)$ at small $\tau$, which is less susceptible to MC error.

\begin{figure}
\includegraphics{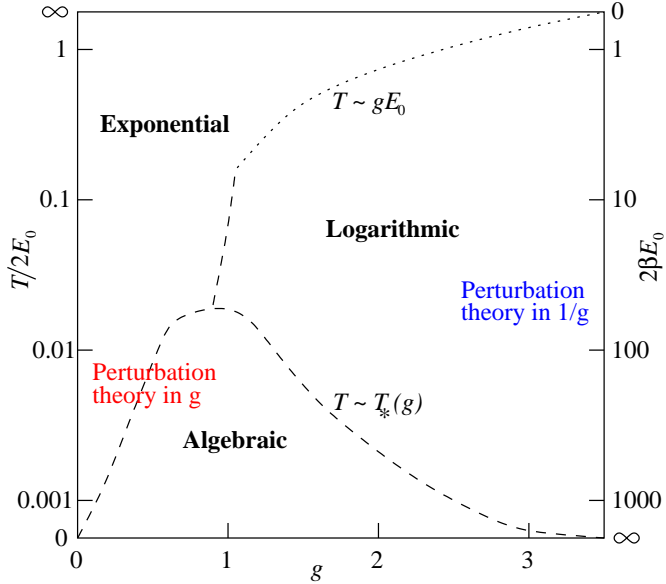}
\caption{{\small \label{f:phasediagram}  
Phase diagram for behavior of $C(\beta/2),\,\beta=1/T$. The algebraic region,
$C(\beta/2)=(T/T_{*})^2$, occurs at a low enough temperature for any
nonzero value of $g$. This phase is dominated by spin-flip
processes. Droplet fluctuations are frozen in the logarithmic phase;
$C(\beta/2)$ decreases slowly as $1-(1/\pi g)\ln(gE_{0}/T)$. As one
approaches criticality (i.e. $g$ increases), spin-flip processes occur
below exponentially low temperatures $T<T_{*}= E_0 / \sqrt{2(e^{\pi g} - 1)}$.   
}}
\end{figure}

Figure \ref{f:phasediagram} is a phase diagram based on the behavior of $C(\beta/2)$ as a function of $g$ and $T$.  
%Since we are dealing with a 1D field theory, 
The dashed and dotted lines represent smooth crossovers 
rather than phase transitions.  Only the $g>1$ part of the phase diagram is relevant to the current paper.  
The crossover between logarithmic and algebraic behavior occurs at $T^* = E_0 / \sqrt{2(e^{\pi g} - 1)}$.

%==============================================================================
\section{Discussion}\label{s:discussion}

We have studied the paramagnon contribution to the spin correlation function
$C(\tau)$ and the static impurity susceptibility $\chi_{dc}(T)$
of a strongly damped $(g>1)$ magnetic defect with $XY$ rotational
symmetry in a metal close to a ferromagnetic quantum critical point.
Our analysis shows that quantum tunneling (droplet's spin flip) effects
are negligible above an exponentially small temperature, 
$T_{*}=E_{0}/\sqrt{2(e^{\pi g}-1)}.$
In this regime, $C(\tau)$ and $T\chi_{dc}(T)$ both
decrease logarithmically as $T$ decreases.  At very low temperatures
$T\ll T_{*},$ $C(\tau)$ is very well described by a power law,
$C(\tau)=(T/T_{*})^{2}/\sin^{2}(\pi T\tau)$,
somewhat analogous to earlier works\cite{chakravarty,ludwig} on
the Coulomb gas model and equivalent Kondo formulations.
The impurity susceptibility $\chi_{dc}(T)$ saturates to an exponentially
large but finite value at $T=0$ and does not show any anomalous divergence.
Near $T=0$, $\chi_{dc}(T)=\chi_{dc}(0)[1-(2/3)(T/T_{*})^{2}],$ 
demonstrating the critical nature of the Ohmic dissipation. While our
results were obtained for an $XY$ defect, we believe they should
be relevant to defects with full spherical symmetry (possibly with additional
modifications due to the Berry phase). We stress that there is a qualitative difference
between the cases of Ising\cite{yuval} and $XY$ symmetry due to the critical nature of
the long-ranged $1/\tau^2$ interaction for one-dimensional 
(meaning one-time dimension) ferromagnetic chains.\cite{kosterlitz}

We considered in this paper the action of Eq.(\ref{millis1}) 
and associated non-perturbative
effects associated with spin flips. These effects are dominant
if, due to the large ferromagnetic polarizability of the host metal,
the corrections proportional to $g$ (corresponding to the even powers of 
exchange coupling $J$) are larger than other Kondo corrections proportional
to odd powers of $Jn(\epsilon_F)$. 
As analyzed by Larkin and Melnikov\cite{larkin}
the condition to neglect odd-power Kondo corrections is $g >
Jn(\epsilon_F)$. Since we consider $g>1$, it is sufficient to require just 
$Jn(\epsilon_F) > (1+F_a)^{1/2}$, 
while the above condition $g > Jn(\epsilon_F)$ is 
satisfied trivially for $ Jn(\epsilon_F)\ll 1$.

The weak coupling case $(g \ll 1)$ is analogous to the overcompensated
Kondo problem studied by Nozieres and Blandin,\cite{nozieres} and
Abrikosov and Migdal\cite{abrikosov} where the
number of conduction electron channels $N_{ch}$ coupling to the defect
is much larger than $2S$. Each conduction electron channel corresponds
to a different orbital quantum number. %   $l$. 
The coupling constant $(J n_F)$ obeys the scaling equation
\begin{equation}
\frac{d (Jn_F)}{d \ln D} = -(Jn_F)^{2} + N_{ch}(Jn_F)^{3} + 
c(Jn_F)^{4}\cdots,   
\label{nozieresrg}
\end{equation}
where $D$ is the bandwidth, and $c$ depends on $S$ but not on
$N_{ch}$. The first term in Eq.(\ref{nozieresrg}) depends on the sign
of the exchange interaction $J$, and for $J>0$ gives rise to the conventional antiferromagnetic
 Kondo effect with a Kondo temperature $T_{K}=D \exp(-1/Jn_F)$. The second
term describes the scattering of a conduction electron from 
the impurity dressed by a number $N_{ch}$ of 
closed electron loops. 
If the second
term is much larger than the first, the coupling constant renormalizes
towards zero as follows:
\begin{equation}
J(D) = \frac{J_{0}}{\sqrt{1 +
    \frac{N_{ch}(J_{0}n_F)^{2}}{2}\ln(D_{0}/D)}}.
\end{equation}
The above scaling is the same as the renormalization of 
$g = (\pi/2)(SJ n_F)^{2}/(1+F_{a})$ 
obtained by Larkin and Melnikov\cite{larkin} for $g \ll 1$ by summing over parquet
diagrams:
\begin{equation}
g(T) = \frac{g}{1 + \frac{2g}{\pi S^{2}} \ln(E_{0}/T)}.
\label{LMrg}
\end{equation}
A comparison of Eq.(\ref{nozieresrg}) and Eq.(\ref{LMrg}) shows
that the number of channels is $N_{ch}=2/(1+F_{a})$.   
As the running coupling constant decreases,
the usual first Kondo term in Eq.(\ref{nozieresrg}) may no longer be disregarded.
If $J>0$ (antiferromagnetic), 
$J(D)$ flows towards a stable multi-channel Kondo fixed point 
given by
\begin{equation} 
J_{*} n_F = 1/N_{ch} = (1+F_{a})/2,
\label{nozieresrg2}
\end{equation}
which corresponds to a Kondo temperature
\begin{equation}
T_{K}=E_{0} e^{-1/J_{*} n_F} = E_0 e^{-2/(1+F_{a})}.
\label{kondotemperature}
\end{equation} 
If $J<0$ (ferromagnetic), the coupling constant scales to zero. In
that case, as in the ferromagnetic Kondo problem,
the impurity's susceptibility is expected to obey a Curie-Weiss law
as $T\rightarrow 0$.

In this paper we have considered $g>1$, and perturbation theory in
$1/g$ also shows that $g$ flows towards zero,\cite{others} 
as is evident from Eq.(\ref{Grenorm}).  Thus, in both the weak-coupling ($g<1$) and the strong-coupling ($g>1$) cases, $g$ is renormalized towards zero until very low temperatures below  $T_{K}=E_{0}\exp[-2/(1+F_{a})]$, when odd-power terms of the coupling $J$ become relevant. %the sign of $J$ becomes relevant. 
%The same conclusion can be made from the renormalization group treatment
%of the overcompensated Kondo problem with number of channels
%$N_{ch}=1/(1+F_a)$ much larger than $2S$.\cite{nozieres} 

The overall picture emerging from our analysis is that the compensation of
the droplet's moment takes place in two basic stages. 
In the first stage, on which we focus in the main text, the paramagnon fluctuations,
which originally enhance a spin $S$ to the large magnetic moment $M$,
are gradually stripped off.
%renormalize the large gyromagnetic ratio of the droplet to the bare value, $S$. 
This quenching of the droplet moment takes place below the
temperature $T_{*}$, which in our case is given by $T_{*}\approx
E_{0}\exp[-\pi g/2]=E_{0}\exp[-(\pi S Jn_F/2)^{2}/(1+F_{a})].$ At the second stage, and
an even lower temperature $T_{K}$, for $J>0$ the usual Kondo effect 
compensates the remaining moment leading to a multi-channel Kondo
fixed point, while for $J<0$, the impurity spin becomes free.  
The multi-channel Kondo fixed point could be an artifact of our
assuming equal coupling of the droplet to all the $N_{ch}$ angular
momentum channels of the conduction electrons. In reality, the
coupling for each channel may be different, and it is possible that
the impurity spin will get compensated successively in different channels. Maebashi, Miyake,
and Varma\cite{varma} recently analyzed the problem in the Kondo
regime and concluded from the scaling equations that the coupling
constants approach a multi-channel Kondo fixed point. They also suggested that
at unattainably low temperatures a crossover happens when a single channel wins out. 
In the last stages of preparation and submission of our manuscript
we became aware of another work\cite{vojta} which arrived to some 
conclusions similar to ours in the case of $O(N)$ spin symmetry in the large-$N$ limit.

%At the temperature $T_K=E_0 \exp[-1/(1+F_a)]$ and below, 
%odd powers of $J$ become relevant,
%and the gyromagnetic ratio of the impurity spin decreases to the bare value.
%For temperatures below $T_K$ the behavior of the impurity depends on the sign
%of the impurity exchange constant $J$. For a ferromagnetic $J<0$,
%free-spin decoupling
%from conduction electrons occurs, while for the antiferromagnetic case $J>0$,
%the impurity spin would get completely compensated.
%Note that the droplet's crossover temperature 
%$T_{*} \sim \exp[-\pi (Jn(\epsilon_F))^2/(2(1+F_a))]$ of spin flips
%is always much larger than the overcompensated multi-channel 
%Kondo temperature $T_K \sim \exp[-1/(1+F_a)]$. 
%Therefore the overall picture is that at temperatures above 
%and below $T_{*}$ the impurity spin is locked into the droplet and 
%behaves as we described above in the paper,
%while at much lower temperatures $T<T_K$,
%for ferromagnetic or antiferromagnetic exchange couplings, 
%the impurity spin can be described by the corresponding, 
%free or completely quenched spin, Kondo effect.

In this paper we considered a dilute system of impurities, so that
mutual interaction among impurities can be neglected. The size of a magnetic
droplet, $L$, is determined by the dispersion relation of the paramagnon
modes and the proximity to the critical point,
as $L \sim \xi_{0}/\sqrt{1+F_a}$. As the quantum critical
point is approached, the size of the droplets grows, and the system
can be considered dilute only if the density of impurities is 
$n_{imp} \ll \xi_{0}^{-3}(1+F_a)^{3/2}$.
We also ignore various anomalous effects which possibly arise
in very close proximity to the critical point,\cite{belitz} and 
explore only the nearly ferromagnetic Fermi-liquid regime.

There are several experimental systems, involving impurities with giant magnetic moments in a nearly ferromagnetic host metal, in which it may be possible to study magnetic droplet phenomena systematically.\cite{korenblit} 
Among these 
are iron ($Fe$) dissolved in various transition metal
alloys,\cite{clogston} nickel ($Ni$) impurities in
palladium\cite{shaltiel,loram1} ($Pd$), and  cobalt ($Co$) impurities in a 
platinum ($Pt$) host.\cite{shen}
% Palladium doped with iron, a
%giant moment alloy frequently used as a magnetic thermometer for low
%temperatures from the sub-millikelvin region to $500mK$ on account of
%a quite accurate Curie-Weiss like susceptibility it shows in this
%temperature range.\cite{jutzler} In designing such alloy thermometers, 
%one needs to stay well away from the crossover temperature $T_{*}$
%where significant deviations from the Curie-Weiss law are expected.    

There are some close connections between the dynamics of a magnetic
moment with $XY$ symmetry and the dynamics of the electromagnetic phase in quantum dots and granular metals.
%Our work is also of relevance to quantum tunneling and Coulomb
%blockade effects in quantum dots and granular metals. 
The power law
behavior of $C(\tau)$ at long-$\tau$ (or small temperatures) can be associated
with  inelastic cotunneling in the literature of mesoscopic physics.\cite{averin}
Inelastic cotunneling was long understood to be 
important at low temperature $(T < E_{0}\ln(g^{-1}))$ 
for weak inter-grain coupling $(g<1)$. 
Our present work shows that this is so even when inter-grain
coupling is large, the difference being that when $g>1$, inelastic cotunneling
becomes important only at exponentially low temperatures
$T<T_{*}$. The competition of inelastic cotunneling and the Coulomb
blockade can lead to interesting consequences for the transport
properties of a granular metal.\cite{loh}

\begin{acknowledgments}
We thank A.J. Millis who brought this problem to our attention, and
P.B. Littlewood, G.G. Lonzarich, J.W. Loram, T.V. Ramakrishnan, and I. Smolyarenko, for 
stimulating discussions. Y.L.L. and V.T. thank Trinity College, Cambridge, for support. 
\end{acknowledgments}

\end{document}